# Continuous User Authentication Using Machine Learning and Multi-Finger Mobile Touch Dynamics with a Novel Dataset

Zach DeRidder
*Deparment of Computer Science*
*University of Wisconsin – Eau Claire*
Eau Claire, US
deriddzt2002@uwec.edu

Nyle Siddiqui
*Deparment of Computer Science*
*University of Wisconsin – Eau Claire*
Eau Claire, US
siddiquin8701@uwec.edu

Thomas Reither
*Department of Computer Science*
*University of Wisconsin – Eau Claire*
Eau Claire, US
reithert2424@uwec.edu

Rushit Dave
*Department of Computer Information Science*
*Minnesota State University, Mankato*
Mankato, US
rushit.dave@mnsu.edu

Brendan Pelto
*Department of Computer Science*
*University of Wisconsin – Eau Claire*
Eau Claire, US
peltobr1555@uwec.edu

Mounika Vanamala
*Department of Computer Science*
*University of Wisconsin – Eau Claire*
Eau Claire, US
vanamalm@uwec.edu

Naaem Seliya
*Department of Computer Science*
*University of Wisconsin – Eau Claire*
Eau Claire, US
seliyana@uwec.edu

*Abstract*—As technology grows and evolves rapidly, it is increasingly clear that mobile devices are more commonly used for sensitive matters than ever before. A need to authenticate users continuously is sought after as a single-factor or multi-factor authentication may only initially validate a user, which doesn't help if an impostor can bypass this initial validation. The field of touch dynamics emerges as a clear way to non-intrusively collect data about a user and their behaviors in order to develop and make imperative security-related decisions in real time. In this paper we present a novel dataset consisting of tracking 25 users playing two mobile games -- Snake.io and Minecraft -- each for 10 minutes, along with their relevant gesture data. From this data, we ran machine learning binary classifiers – namely Random Forest and K-Nearest Neighbor - to attempt to authenticate whether a sample of a particular user's actions were genuine. Our strongest model returned an average accuracy of roughly 93% for both games, showing touch dynamics can differentiate users effectively and is a feasible consideration for authentication schemes. Our dataset can be observed at https://github.com/zderidder/MC-Snake-Results

## I. INTRODUCTION

In the past 5 to 10 years, an ever-increasing reliance on technology to assist in corporate and personal matters including handling private employee information, accessing confidential data relating to government and tax functions, as well as utilizing online banking applications for financial reasons has become realized. Formerly, many of these applications were largely limited to resources like laptops and personal computers. However, technological revolution and a stronger desire to have access to this information at the touch of the fingertips has been actualized in recent years. In between 2016 and 2022, the number of mobile smartphone subscriptions has increased from 3.668 billion to 6.567 billion, which is an increase of 79% [1]. Mobile devices often get lost or are stolen and be corrupted with no form of protection against attackers if tracking or location services are manually turned off [2-4]. Touch dynamics could be enabled on these devices and offer a final level of protection against somebody who could gain unauthorized access to these devices. This rapid growth in accessibility to devices with important personal data illustrates a need to develop more secure authentication schemes on these devices [5, 6].

How users interact with their mobile devices and the various features that can be extracted from their touch data is generally known as touch dynamics. Touch dynamics can be categorized into two main subcategories for the purposes of user authentication, with those areas being keystroke based and gesture-based touch dynamics [7, 8]. A keystroke approach focuses most on how users interact with typing out characters or words on a mobile keyboard via individual taps. This differs from a gesture-based approach which would focus more on an aggregate of data points in continuous motion, commonly referred to as a swipe or gesture. Much of the existing gesture-based literature however commonly focuses on one handed gestures as they are the most common way to interact with a mobile device [7]. In fact, about half of mobile device users prefer to use one hand when interacting with devices [9].

Therefore, our research was driven by the following contributions to the academic sphere:
- Develop various binary classifiers (RF, KNN) to evaluate the efficacy of multi-finger touch dynamics and compare the classifier results
- Contribute a publicly available dataset of 25 users playing 2 mobile games, Snake.io and Minecraft, to the academic sphere. This dataset is available for download at https://github.com/zderidder/MC-Snake-Results
- Comparing the efficiency of binary classifiers in a multi-finger environment in comparison to previous works which heavily focus on single-finger gesture-based datasets.

## II. BACKGROUND AND RELATED WORK

Touch dynamics as a field has been growing more popular in the eyes of researchers in the past 10 years and doesn't appear to be slowing down any time soon. As recently as 2016, a survey paper on touch dynamics in mobile devices noted that there were only 3 publicly available datasets at the time of writing in relation to this field [10]. In a mere 3 years after [10], [11] already had 11 datasets referenced to compare a varying number of subjects, input lengths of strings, and multitudes of other datapoints. However, a vast majority of these datasets are all belonging to the keystroke-based touch dynamics subset referenced in the introduction. Therefore, we opted to primarily track biometric data in a non-keyboard environment on mobile devices in order to contribute more data to the gesture-based side of touch dynamics. More of the information relating to this gestured-based feature extraction is described in Section III.

An effective biometric system can be hard to boil down to exact parameters in terms of what to focus on and utilize for researchers. [12-14] as a collection helps to set a strong foundation for defining effective characteristics of a biometric system, including a focus on the unobtrusiveness to the user experience, transparency for what is being collected on the device, and most importantly the ability to continually observe and make decisions based on user's gestures. While we do not make decisions based on user's data in real time, we set out to analyze what is possible for differentiating users based on multiple simultaneous inputs.

Various touch dynamics studies have previously opted to continuously authenticate users using several machine learning models. One paper aimed to study general phone usage from a sample size of 41 users, tracking phone orientation, x and y coordinates, and pressure [15]. From their findings, they noted that the various pressure sensors were the most significant features in authenticating users. They concluded that authentication of users is feasible with a highest-score accuracy of 83.60% based solely on interactions with their mobile phones [15].

[16] also sets a good foundation for analyzing touch gestures involving multiple fingers and provides additional touch information utilizing sensors from a touch glove. They analyzed normal daily usage of a phone continuously from 40 users who were wearing a glove that has pressure sensors present on two different fingers simultaneously. Interestingly, they found that the false-acceptance rates and false-rejection rates were far lower with additional glove touch sensor information. Ultimately, they were able to conclude that touch gestures from different people on mobile devices can be used as a source of information for user authentication.

A sizeable motivation for this paper aims to build on a previous mouse dynamics work focusing on high-intensity novel data collection, with our focus being into the field of mobile touch dynamics with dual sets of datapoints. The processes in question can be found in two papers [17, 18], which aimed to authenticate different users who had played the computer game *Minecraft* solely based on mouse movement data via binary classifiers.

## III. METHODOLOGIES

### A. Data Collection

Our dataset collected 25 user's touch data while they played both *Snake.io* and *Minecraft* on mobile phones for 10 minutes each. All of the settings and configurations of both the games were standardized in order to ensure that all users played the game in the same environment. These games were chosen since they require two fingers to be used simultaneously at opposing ends of the screen, leading to more data collected and more potential variability among subjects. Two different Android phones were used in data collection, one being a Samsung Galaxy S10E and the other being an LG V30+. The phones were held in a landscape position by all users during data collection.

In order to collect the data into a meaningful format, a Python script was developed to connect to the Android phones sensors via the Android debugging interface and an algorithm was derived to differentiate between the fingers from the raw sensor data. The raw data of all users was then saved to a comma separated value text file, where each line of data represented a single touch event per finger. Each line of text included 7 fields: (*Timestamp, X, Y, Button Touch, Touch Major, Touch Minor, Finger*) where the Timestamp is the time represented in seconds, X and Y are the x-coordinate and y-coordinate of the device, Button Touch is whether a finger is currently being held on the screen, Touch Major and Touch Minor represent the length of the ellipse of the touch point on the screen at the point of contact on the major axis and minor axis, respectively. The finger column simply refers to which of the two fingers the data is being attributed to, with possible values of 0 or 1 for each finger.

We also strived for a sensible number for grouping of touch datapoints so that they can contain enough data to be used later for feature extraction with enough variability among subjects. We chose to avoid the route of using a time-based approach to define gestures as there is high variability in a specific time frame, while a fixed number can make all gestures equal in data represented [19]. We chose to group events like [17], noting that 10 separate tapping/sliding events makes for meaningful data in the feature extraction and is sufficient to train the model later.

| | A | B | C | D | E | F | G | H | I |
|---|---|---|---|---|---|---|---|---|---|
| 1 | Timestamp | X | Y | BTN_TOUCH | TOUCH_MAJOR | TOUCH_MINOR | FINGER | | |
| 5352 | 55.021863 | 1721 | 458 | HELD | 27 | 19 | 0 | | |
| 5353 | 55.030851 | 1723 | 461 | HELD | 26 | 18 | 0 | | |
| 5354 | 55.030851 | 1156 | 3612 | UP | 15 | 11 | 1 | | 10 datapoints which represent 1 gesture |
| 5355 | 55.03895 | 1723 | 463 | HELD | 26 | 18 | 0 | | |
| 5356 | 55.03895 | 1156 | 3614 | HELD | 10 | 10 | 1 | | |
| 5357 | 55.04704 | 1722 | 465 | HELD | 26 | 18 | 0 | | |
| 5358 | 55.04704 | 1156 | 3618 | HELD | 11 | 11 | 1 | | |
| 5359 | 55.055571 | 1715 | 466 | HELD | 26 | 18 | 0 | | |
| 5360 | 55.055571 | 1156 | 3621 | HELD | 13 | 13 | 1 | | |
| 5361 | 55.063779 | 1710 | 467 | HELD | 25 | 17 | 0 | | |

Fig 1. An illustration of what a gesture could look like. In this example, this is not entirely accurate as the finger data is not separated, but this gives a general idea of how it can be represented. Multi-finger data points highlighted in red to show similar timestamps.

### B. Data Cleaning

Before being able to extract features from our data, mandatory steps were taken to avoid potential errors in our methodologies. Firstly, rows with null values were omitted from the data for all 25 users. It was also made sure that the various types of information and device actions were consistent between the two different types of phones. Another issue arose with multiple events happening at the same time. Due to the nature of the Android debugging interface and its output stream, two events could happen at the same timestamp from each finger, as can be seen in Figure 1.

This led to issues in some functions with dividing by zero errors, so the data from each finger was sorted into its own section, with all of the data from finger 1 first, then all of the data from finger 2 at the bottom. This also helped to make sure that the calculations for each finger did not interfere with the other in the section where we discuss feature extraction. Later in the process, all the separated finger data would be combined randomized to include data from each finger at random intervals when the classifiers would pick gesture data to analyze.

### C. Feature Extraction

From the data we collected for each user there were 9 other features that were extracted from the original fields. These features are necessary since they are used to identify users since each feature is distinctive to each user [20]. These 9 features include: X Speed, Y Speed, Speed, X acceleration, Y Acceleration, Acceleration, Jerk, Path Tangent, Angular Velocity, and each feature is defined in Table 1 below.

| Feature Name | Equation |
|---|---|
| X speed | $X_i - X_{i-1}/T_i - T_{i-1}$, with X being X position and T being time |
| Y Speed | $Y_i - Y_{i-1}/T_i - T_{i-1}$, with Y being Y position and T being time |
| Speed | sqrt (XSpeed^2 + YSpeed^2), where sqrt is a square root function and XSpeed and YSpeed are from the features described above. |
| X accel. | $X_i - X_{i-1}/T_i - T_{i-1}$, with X being X speed and T being time |
| Y accel. | $Y_i - Y_{i-1}/T_i - T_{i-1}$, with Y being Y speed and T being time |
| Accel. | $S_i - S_{i-1}/T_i - T_{i-1}$, with S being speed values from the equation above, and T being time |
| Jerk | $A_i - A_{i-1}/T_i - T_{i-1}$, with A being acceleration and T being time |
| Path Tan. | arctan2($Y_i - Y_{i-1}/X_i - X_{i-1}$), with Y being the Y position, X being the X position, and the arctangent2 function |
| Angular Velocity | $P_i - P_{i-1}/T_i - T_{i-1}$, with P being the path tangent function defined above and T being time |

Table 1. The various features extracted and their representative equations.

All the features above in Table 1 are vital in giving the binary classifiers enough variance to differentiate users based on gesture data. From these 9 features, the average, standard deviation, minimum value and maximum value for the 10 datapoints in each gesture were also collected, for a total of 36 features. These biometrics features are of the most importance as behavioral biometrics and related features are extremely personal to each user [21]. Not only are behavioral biometrics different for each user but psychological features among users can create notable differences as well [22].

### D. Training and Testing

For the training and testing of our classifiers, the total data generated from each user was split into 80% being used for training and 20% being used for testing. Most papers will tend to range from 70% to 80% in training data, and we decided to go with 80% as this is empirically found to garner the strongest results [23]. The training and testing data for all users was then concatenated and shuffled into a master file, to avoid bias and any sense of "order" for the classifiers. When splitting the data for processing, the authentic user's data took up 50% and the other 50% came equally from all the other imposters data in order to mitigate bias. In order to determine whether data was authentic or not among the training data, each point was given a value of 0 or 1, with 0 being authentic and 1 being imposter data. These values were then checked while the testing data by running them against the model to determine accuracy of the classifier.

## IV. RESULTS AND ANALYSIS

Our paper aims to evaluate our novel dataset by attempting to differentiate between users simply based on the features extracted from their gesture interactions with the mobile devices presented. Our main method of analysis utilizes binary classifiers, namely Random Forest (RF) and K-nearest Neighbor (KNN). Since the data was equally sourced from all users in Section III, all the classifiers were trained and tested on individual users, either from their time playing Minecraft or Snake.io.

In order to analyze the performance of each binary classifier, the following evaluation components were present in the results for training on each individual user out of 25:

- Accuracy, the overall percentage of correctly identified results to inaccurate results
- False Positive Rate (FPR), which falsely authenticates an impostor
- False Negative Rate (FNR), which falsely denies a genuine user access
- True Positive Rate (TPR), which correctly identifies the genuine user
- True Negative Rate (TNR), which correctly and denies the impostor class
- Equal Error Rate (EER), which expresses a threshold where FPR and FNR are equal.

Of these various evaluation metrics, it is preferential to have the lowest possible scores for FPR and FNR as in a real-world evaluation of a biometric system, FPR could authorize an impostor to access confidential information when they should be denied access. Accuracy is also particularly useful as it can show the overall performance authenticating a particular user in relation to their testing data. EER is similarly useful in showing the overall performance and accuracy of a classifier.

### A. Snake.io results

| User | Game | Model | Accuracy | FPR | FNR | EER |
|---|---|---|---|---|---|---|
| 1 | Snake | RF | 0.9910 | 0.0042 | 0.0138 | 0.0090 |
| 2 | Snake | RF | 0.9158 | 0.0138 | 0.1766 | 0.0952 |
| 3 | Snake | RF | 0.9063 | 0.0183 | 0.1941 | 0.1062 |
| 4 | Snake | RF | 0.9341 | 0.0146 | 0.1282 | 0.0714 |
| 5 | Snake | RF | 0.9181 | 0.0245 | 0.1692 | 0.0919 |
| 6 | Snake | RF | 0.9158 | 0.0253 | 0.1575 | 0.0914 |
| 7 | Snake | RF | 0.9153 | 0.0109 | 0.1832 | 0.0971 |
| 8 | Snake | RF | 0.9193 | 0.0142 | 0.1667 | 0.0905 |
| 9 | Snake | RF | 0.9117 | 0.0135 | 0.1883 | 0.1009 |
| Avg | Snake | RF | 0.9339 | 0.0471 | 0.1355 | 0.0913 |
| Avg | Snake | KNN | 0.834 | 0.0635 | 0.2964 | 0.2100 |
| Stdv | Snake | RF | 0.0203 | 0.1794 | 0.0431 | 0.0924 |
| Stdv | Snake | KNN | 0.0450 | 0.0245 | 0.0947 | 0.0783 |

Table 2 – a fraction of the user results from the Random Forest model and the overall average results for each evaluation metric. Key: Avg = average, Stdv = Standard Deviation. Full values for all users are omitted to conserve space, but can be seen at https://github.com/zderidder/MC-Snake-Results

The classifiers returned mostly similar results across both sets of games. The average accuracy for the RF turned out to be 93.39% while KNN ended up with an accuracy of 83.4%. FPR and FNR are preferential to have as low as possible, as those could be inaccurate judgments that could lead to security breaches if access was granted to an impostor. In case of FPR, this is very good as the average is 4.71% for RF, alongside 6.35% for KNN. FNR is a bit higher in both instances at 13.54% for RF and 29.64% for KNN. The dataset for Snake.io contains a cumulative 1,586,684 gesture events. While it is hard to compare to previous works due to the simultaneous multi-finger approach to our paper, some previous works have achieved accuracies of 89% with a repeated, single-finger gesture-based approach [24]. Similar findings have been found in papers that capture normal phone data usage, finding accuracies of 80-85% depending on the model [25]. Due to a larger sample size of users and more variability in types of actions in our paper, this appears to be a viable application for high-intensity, randomized movements.

### B. Minecraft results

| User | Game | Model | Accuracy | FPR | FNR | EER |
|---|---|---|---|---|---|---|
| 1 | MC | RF | 0.9784 | 0.0076 | 0.0364 | 0.0220 |
| 2 | MC | RF | 0.9174 | 0.0139 | 0.1722 | 0.0931 |
| 3 | MC | RF | 0.9195 | 0.0198 | 0.1569 | 0.0884 |
| 4 | MC | RF | 0.9195 | 0.0117 | 0.1705 | 0.0911 |
| 5 | MC | RF | 0.9130 | 0.0195 | 0.1743 | 0.0969 |
| 6 | MC | RF | 0.9239 | 0.0233 | 0.1406 | 0.0820 |
| 7 | MC | RF | 0.9135 | 0.0108 | 0.1882 | 0.0995 |
| 8 | MC | RF | 0.9112 | 0.0143 | 0.1882 | 0.1013 |
| 9 | MC | RF | 0.9255 | 0.0156 | 0.1484 | 0.0820 |
| Avg | MC | RF | 0.9350 | 0.0133 | 0.1297 | 0.0715 |
| Avg | MC | KNN | 0.8329 | 0.0690 | 0.3037 | 0.2072 |
| Stdv | MC | RF | 0.1867 | 0.0050 | 0.0421 | 0.0221 |
| Stdv | MC | KNN | 0.0261 | 0.0136 | 0.0582 | 0.0522 |

Table 3 – some of the user results from the Random Forest model and the overall average results for each evaluation metric for each model. Key: Avg = average, Stdv = Standard Deviation. See full results in link in Table 2.

Similar to the results for Snake.io, the Minecraft results had an average accuracy of 93.49% for the RF classifier while the KNN ended up with an average accuracy of 83.29%. The FPR in Minecraft was on average, 1.33% for RF and 6.89% for KNN, which likely can be attributed to more variability in the gameplay and less data to make decisions on, since Snake.io was more intense for the 10 minute fixed time period. The FNR for Minecraft was 12.97% for RF and 30.37% for KNN as well. Collectively, the Minecraft dataset contains data for 866,233 gesture events.

### V. LIMITATIONS AND FUTURE WORK

The use of multi-finger authentication on mobile devices can offer more precision due to more datapoints, yet there are a few limitations that could make it less practical in a real setting. One drawback is that it will require two finger inputs continuously in order to make its predictions accurately. It is far more common for users to interact with mobile devices with only one finger, which might narrow use cases for this application. Another requirement for authenticating users is that there must be a dataset linked to a user before they are able to be authenticated without adaptive technologies.

Without knowing a user's profile and mannerisms prior to authentication, there will be no way of knowing whether the user is authentic since there is no previous information to base the predictions off. This can be resolved by having users calibrate their own touch dynamics data prior to authentication like facial recognition on a device such as an *iPhone* or *Android*. For the future we plan to use different types of machine learning classifiers in order to gain a more comparative study on all different types of techniques. We also plan to improve on being able to detect an imposter without prior user data. The classifiers would then justify its predictions based on the data already acquired and use that to determine if this new information matches up with any preexisting data.

## VI. CONCLUSION

Throughout this paper we introduced a new mobile touch dynamics dataset with the use of simultaneous, multi finger events while playing the mobile games Snake.io and Minecraft. This data collection was carried out with the intention of contributing knowledge to the field of continuous user authentication with mobile touch dynamics, while also providing a unique dataset to the field for others to similarly reference and analyze. We have shown that in our two 10-minute gaming sessions that enough data was collected for the machine learning models to have average accuracies of roughly 93% with RF and 83% for KNN in differentiating between authentic users and impostors.

Our findings propose that tracking multiple events concurrently on a given device could reasonably lead to authenticating users using a non-intrusive, passively running authentication system, given enough time to collect user data to make a unique profile. Going forward, we hope these findings help to formulate a way to dynamically track users and make authentication decisions in real-time.

## REFERENCES

[1] O'Dea, S. (2022, February 23). Number of smartphone subscriptions worldwide from 2016 to 2027. Retrieved from https://www.statista.com/statistics/330695/number-of-smartphone-users-worldwide/

[2] Budulan, Ş., Burceanu, E., Rebedea, T., Chiru, C. (2015). Continuous User Authentication Using Machine Learning on Touch Dynamics. In: Arik, S., Huang, T., Lai, W., Liu, Q. (eds) Neural Information Processing. ICONIP 2015. Lecture Notes in Computer Science(), vol 9489. Springer, Cham. https://doi.org/10.1007/978-3-319-26532-2_65

[3] J. Shelton *et al.*, "Palm Print Authentication on a Cloud Platform," *2018 International Conference on Advances in Big Data, Computing and Data Communication Systems (icABCD)*, 2018, pp. 1-6, doi: 10.1109/ICABCD.2018.8465479.

[4] D. J. Gunn, Z. Liu, R. Dave, X. Yuan, and K. Roy, "Touch-based Active Cloud Authentication Using Traditional Machine Learning and LSTM on a distributed Tensorflow framework," International Journal of Computational Intelligence and Applications, vol. 18, no. 04, p. 1950022, 2019.

[5] Ackerson JM, Dave R, Seliya N. Applications of Recurrent Neural Network for Biometric Authentication & Anomaly Detection. *Information*. 2021; 12(7):272. https://doi.org/10.3390/info12070272

[6] Strecker, Sam, et al. "A Modern Analysis of Aging Machine Learning Based IoT Cybersecurity Methods." arXiv preprint arXiv:2110.07832 (2021).

[7] Wang, K., Zhou, L., Zhang, D., Liu, Z., & Lim, J. (2020). What is More Important for Touch Dynamics based Mobile User Authentication?\

[8] Dave, R., Seliya, N., Pryor, L., Vanamala, M., & Sowells, E. (2022). Hold On and Swipe: A Touch-Movement Based Continuous Authentication Schema based on Machine Learning. arXiv preprint arXiv:2201.08564.

[9] Wang, Kanlun & Zhou, Lina & Zhang, Dongsong. (2019). User Preferences and Situational Needs of Mobile User Authentication Methods. 18-23. 10.1109/ISI.2019.8823274.

[10] Teh, Pin Shen & Zhang, Ning & Tan, Syh-Yuan & Shi, Qi & How, Khoh & Nawaz, Raheel. (2020). Strengthen user authentication on mobile devices by using user's touch dynamics pattern. Journal of Ambient Intelligence and Humanized Computing. 11. 10.1007/s12652-019-01654-y.

[11] Teh, P. S., Zhang, N., Teoh, A. B. J., & Chen, K. (2016). A Survey on Touch Dynamics Authentication in Mobile Devices. Computers and Security.

[12] Alghamdi, Shatha & Elrefaei, Lamiaa. (2017). Dynamic Authentication of Smartphone Users Based on Touchscreen Gestures. Arabian Journal for Science and Engineering. 43. 10.1007/s13369-017-2758-x.

[13] Siddiqui N., Pryor L., Dave R. (2021) User Authentication Schemes Using Machine Learning Methods—A Review. In: Kumar S., Purohit S.D., Hiranwal S., Prasad M. (eds) Proceedings of International Conference on Communication and Computational Technologies. Algorithms for Intelligent Systems. Springer, Singapore. https://doi.org/10.1007/978-981-16-3246-4_54

[14] Pryor, L., Dave, D., Seliya, D., Boone, D., & Sowells, E. R. (2021). Machine Learning Algorithms In User Authentication Schemes. arXiv preprint arXiv:2110.07826.

[15] Budulan, Ş., Burceanu, E., Rebedea, T., Chiru, C. (2015). Continuous User Authentication Using Machine Learning on Touch Dynamics. In: Arik, S., Huang, T., Lai, W., Liu, Q. (eds) Neural Information Processing. ICONIP 2015. Lecture Notes in Computer Science(), vol 9489. Springer, Cham. https://doi.org/10.1007/978-3-319-26532-2_65

[16] T. Feng et al., "Continuous mobile authentication using touchscreen gestures," 2012 IEEE Conference on Technologies for Homeland Security (HST), 2012, pp. 451-456, doi: 10.1109/THS.2012.6459891.

[17] N. Siddiqui, R. Dave and N. Seliya, "Continuous User Authentication Using Mouse Dynamics, Machine Learning, and Minecraft," *2021 International Conference on Electrical, Computer and Energy Technologies (ICECET)*, 2021, pp. 1-6, doi: 10.1109/ICECET52533.2021.9698532.

[18] Siddiqui N, Dave R, Vanamala M, Seliya N. Machine and Deep Learning Applications to Mouse Dynamics for Continuous User Authentication. *Machine Learning and Knowledge Extraction*. 2022; 4(2):502-518. https://doi.org/10.3390/make4020023

[19] Meng, Weizhi & Wong, Duncan & Kwok, Lam. (2014). Design of touch dynamics based user authentication with an adaptive mechanism on mobile phones. Proceedings of the ACM Symposium on Applied Computing. 10.1145/2554850.2554931.

[20] Teh, P. S., Chen, K., & Zhang, N. (2019). Using users' touch dynamics biometrics to enhance authentication on mobile devices. University of Manchester. Retrieved from https://www.research.manchester.ac.uk/portal/files/159168194/FULL_TEXT.PDF

[21] Robson de, O. A., Dino, M. A., William, F. G., & Rafael Timóteo de Sousa Júnior. (2021). A framework for continuous authentication based on touch dynamics biometrics for mobile banking applications. Sensors, 21(12), 4212. doi:https://doi.org/10.3390/s21124212

[22] Zhang, X., Zhang, P., & Hu, H. (2021). Multimodal continuous user authentication on mobile devices via interaction patterns. Wireless Communications & Mobile Computing (Online), 2021 doi:https://doi.org/10.1155/2021/5677978

[23] Gholamy, Afshin; Kreinovich, Vladik; and Kosheleva, Olga, "Why 70/30 or 80/20 Relation Between Training and Testing Sets: A Pedagogical Explanation" (2018). Departmental Technical Reports (CS). 1209.

[24] K. W. Nixon, Xiang Chen, Z. -H. Mao and Yiran Chen, "SlowMo - enhancing mobile gesture-based authentication schemes via sampling rate optimization," 2016 21st Asia and South Pacific Design Automation Conference (ASP-DAC), 2016, pp. 462-467, doi: 10.1109/ASPDAC.2016.7428055.

[25] Yuxin Meng, Duncan S. Wong, and Lam-For Kwok. 2014. Design of touch dynamics based user authentication with an adaptive mechanism on mobile phones. In Proceedings of the 29th Annual ACM Symposium on Applied Computing (SAC '14). Association for Computing Machinery, New York, NY, USA, 1680–1687. https://doi.org/10.1145/2554850.2554931